\documentstyle[12pt]{article}
\newcommand{\draft}{
        \renewcommand{\baselinestretch}{1.0}%
        \small\normalsize%
}

\draft
\begin{document}
\title{\bf Acceleration of particles in pulsar magnetosphere and the 
X-ray radiation}
\author{Oktay H. Guseinov$\sp{1,2}$
\thanks{e-mail:huseyin@gursey.gov.tr},
A\c{s}k\i n Ankay$\sp1$
\thanks{e-mail:askin@gursey.gov.tr}, \\
Sevin\c{c} O. Tagieva$\sp3$
\thanks{email:physic@lan.ab.az}, 
\\ \\
{$\sp1$T\"{U}B\.{I}TAK Feza G\"{u}rsey Institute} \\
{81220 \c{C}engelk\"{o}y, \.{I}stanbul, Turkey} \\
{$\sp2$Akdeniz University, Physics Department,} \\ 
{Antalya, Turkey} \\
{$\sp3$Academy of Science, Physics Institute,} \\
{Baku 370143, Azerbaijan Republic} \\
}

\date{}
\maketitle
\begin{abstract}
\noindent
The available data of single X-ray pulsars, their wind nebulae, and the 
SNRs which are connected to some of these sources are analysed. It is 
shown that electric field intensity of neutron stars tears off charged 
particles from the surface of neutron star and triggers the acceleration 
of particles. The charged particles are accelerated mainly in the field of 
magnetodipole radiation wave. Power and energy spectra of the charged 
particles depend on the strength of the magnetodipole radiation. 
Therefore, the X-ray radiation is strongly dependent on the rate of 
rotational energy loss and weakly dependent on the electric field 
intensity. Coulomb interaction between the charged particles is the  
main factor for the energy loss and the X-ray spectra of the charged 
particles.   
\end{abstract}

\section{Introduction}
The number of single X-ray pulsars and the radio pulsars which have been 
detected in X-ray with characteristic time ($\tau$) up to 10$^6$ yr 
located at distances less than 7 kpc from the Sun 
together with 2 pulsars in Magellanic Clouds is 33. These pulsars have 
been observed in 0.1-2.4 keV band (Becker \& Trumper 1997; Becker \& 
Aschenbach 2002) and in 2-10 keV band (Possenti et al. 2002) and all the 
available data have been collected, revised and put together in Table 1. 
All of these pulsars have rate of rotational energy loss  
\.{E}$>$3$\times$10$^{34}$ erg/s. Nine of these 33 pulsars are connected 
to F-type supernova remnants (SNRs) and 8 of them are connected to C-type 
SNRs. Therefore, there exist pulsar wind nebulae (PWNe) near these 17 
pulsars which are located at d$<$7 kpc. Three pulsars are connected to 
S-type SNRs and 3 other pulsars are connected to SNRs for which the 
morphological type is not known. Ten pulsars with detected X-ray radiation 
are not connected to SNRs and 6 of them have characteristic ages 
2$\times$10$^5$$<$$\tau$$<$10$^6$ yr. Among the 33 pulsars, 2 of 
them are X-ray pulsars from which radio radiation has not been observed
and they are connected to SNRs (pulsar J1846-0258 to SNR G29.7-0.3 and 
pulsar J1811-1925 to SNR G11.2-0.3). Radio pulsar J1646-4346 
(d=4.51 kpc) which is connected to C-type SNR G341.2+0.9 but not observed 
in X-ray is also included in the list. Apart from these sources, we have 
also included 14 radio pulsars with \.{E}$>$4$\times$10$^{35}$ erg/s, 
\.{P}$>$10$^{-15}$ s/s and d$\le$7 kpc in Table 1. Also, 6 single 
millisecond pulsars from which X-ray radiation have been observed are 
displayed in Table 1.  

The main aim in this work is to analyse the conditions which are 
necessary for the acceleration of relativistic particles, the X-ray 
radiation produced in the magnetospheres of pulsars and their PWNe.  

\section{Dependence of the ejection of relativistic particles on different 
parameters of pulsars}
As known, very young pulsars have the capability to eject relativistic 
particles. Power and spectra of particles must be related to different 
parameters of pulsars, particulary to the induced electric field  
intensity of which it is practically impossible to find the exact 
expression, because there are large uncertainties in the magnetic field 
structures of neutron stars in plasma. Therefore, we use the simplified 
relation
\begin{equation}
E_{el} = \frac{B_r R}{P}
\end{equation}
(Lipunov 1992). Here, R is the radius, B$_r$ is the real value of 
the magnetic field intensity on the pole, P is the spin period of the 
neutron star and c is the speed of light. Naturally, number and energy 
spectra of relativistic particles also depend on  
the conditions in the atmospheres of neutron stars (density and 
temperature). Independent of the mechanism, the rate of rotational energy 
loss (\.{E}) depends only on the moment of inertia (I), the spin period 
and the time derivative of the spin period (\.{P})
\begin{equation}
\dot{E}=\frac{4\pi ^2 I \dot{P}}{P^3}.
\end{equation} 

If we assume that practically the rate of rotational energy loss is 
connected only to magnetodipole radiation (Lipunov 1992; Lyne \& 
Graham-Smith 1998), then the value of the effective magnetic field 
B=B$_r$ and 
\begin{equation}
\dot{E} = \frac{2}{3} \frac{\mu^2 \omega^4}{c^3} = \frac{32\pi^4}{3c^3} 
\frac{B^2R^6}{P^4} 
\end{equation}
where $\mu$ is the magnetic moment of neutron star.
From expressions (1) and (3) we get 
\begin{equation}
\dot{E} = \frac{2\pi^2}{3c} \frac{R^4}{P^2} E_{el}^2.
\end{equation}
As seen from this equation, \.{E} is directly proportional to E$_{el}$$^2$ 
and inversely proportional to P$^2$. 

From expressions (1)-(3), we can get
\begin{equation}
E_{el} \sim I^{1/2} R^{-2} \tau ^{-1/2}
\end{equation}
from which it follows that the lines of E$_{el}$=constant roughly pass 
parallel to 
the lines of $\tau$=constant on the P-\.{P} diagram (where I and R are 
assumed to be constant). Dependences of $\tau$, E$_{el}$, B, and \.{E} on 
the values of I, R, and the braking index (n) and also the evolutions of 
pulsars with very different parameters are examined in detail by Guseinov 
et al. (2003a).

The existing theories do not give us the possibility to estimate the value 
of the voltage better than the order of magnitude. But it is exactly known 
that charged particles can be pulled out from the hot atmosphere 
of pulsar because of the large value of E$_{el}$ and they can be 
accelerated. Further acceleration can take 
place in the field of the magnetodipole wave (Lipunov 1992). Which of 
these mechanisms is mainly responsible for the acceleration must be 
determined from the observational data. It may be the case that the 
voltage creates only basis for further acceleration of particles similar 
to the case of the Maxwell tail of hot particles (or particles which have 
already been accelerated with some other mechanism) in 
the shock fronts of SNRs. In the regular acceleration models (Bell 1978a, 
1978b, Krymsky 1977), the further acceleration of particles with high 
energy takes place under the crossing through the shock 
fronts of SNRs (Allakhverdiev et al. 1986). Therefore, it is necessary to 
analyse 
the observational data. 

\section{How are the observational data related to the theoretical idea?}
From the observations of the radio and X-ray radiation of neutron stars 
with power-law spectra (excluding the cooling radiation of neutron 
stars), there is direct evidence for continuing particle acceleration. 
Such X-ray radiation have been observed, for example, from pulsars 
J1856+0113 and J1801-2451 which 
have PWNe and have values of \.{E}$\sim$(4-5)$\times$10$^{35}$ erg/s and 
2.5$\times$10$^{36}$ erg/s, respectively. On the other hand, millisecond 
pulsars J1939+2134 and J1824-2452 with \.{E}$\sim$(1-2)$\times$10$^{36}$ 
erg/s also have similar X-ray radiation as the previous pulsars. 
As seen in Table 1, these pulsars, beginning from J1856+0113, have values 
of X-ray radiation 10$^{33}$, 1.5$\times$10$^{33}$, 5$\times$10$^{32}$, 
and 3.6$\times$10$^{33}$ erg/s, respectively, in the 2-10 keV band 
(Possenti et al. 2002; Becker \& Aschenbach 2002). 
We have estimated the ratio  
($\frac{\dot{P}}{P}$)$^{1/2}$$\sim$$\tau$$^{-1/2}$ from the dependence of 
E$_{el}$ (see expression (5)) for pulsars J1856+0113, J1801-2451, 
J1939+2134, and J1824-2452 and we have found the values 
9$\times$10$^{-7}$, 10$^{-6}$, 7$\times$10$^{-9}$, and 3$\times$10$^{-8}$ 
s$^{-1/2}$, respectively. If in all these four cases the values of radius 
and moment of inertia of the pulsars are approximately the same, then the 
values of the electric field intensity may also have approximately the 
same ratios. 

In Figure 1, dependence of the X-ray luminosity in 2-10 keV band 
(L$_{2-10keV}$) on the value of \.{E} for the single pulsars with 
distances up to 5 kpc from the Sun together with 2 pulsars in the 
Magellanic Clouds (see Table 1) is represented. As seen in Figure 1, 
pulsars J1856+0113 and J1801-2451 have \.{E} values  
similar to single millisecond pulsars J1824-2452 and J1939+2134, and 
their \.{E} values are more than 300 times greater than their X-ray 
luminosities in 2-10 keV band. 
Therefore, the luminosity of the power-law X-ray radiation (which  
includes the radiation of the relativistic particles in the Coulomb and 
magnetic fields) mainly depends on the value of \.{E}. How can this 
approach be confirmed also for other pulsars based on the observational 
data? 

Since we are interested in the acceleration of particles, we have 
chosen the pulsars for which 
L$_{2-10keV}$/L$_{0.1-2.4keV}$$>$1, so that, we avoid the errors connected 
to the interstellar absorption and the radiation related with the cooling 
of neutron stars. As seen from Table 1, these conditions are not 
satisfied for the last 4 pulsars with log $\tau$=9.58-9.86 which are also 
displayed in Figure 1. But they are very close to the Sun and very old, 
therefore they do not create any difficulties. All the young pulsars 
shown in Figure 1 have not only hard spectra but also have PWNe. As seen 
from the equation of the best fit
\begin{equation}
L_{2-10keV} = 10^{-19.46 \pm 3.32} \dot{E}^{1.45 \pm 0.09}
\end{equation}
there exists well a dependence between L$_{2-10keV}$ and \.{E} for which 
the radiation is related with the accelerated particles. 

In Figure 2, the relation between L$_{2-10keV}$ and the characteristic 
time of the same pulsars shown in Figure 1 is represented. As seen from 
the figure, a single dependence for both the young and the old pulsar 
populations does not exist (see also Possenti et al. 2002). Pulsars with 
similar values of L$_{2-10keV}$ and \.{E} may have very different 
values of characteristic 
time and this leads to different dependences for each population. As 
their ages increase, the young pulsars continue to move practically along 
the same direction on the L$_x$-$\tau$ diagram. Therefore, the electric 
field intensity does not determine the intensity and maybe also the 
spectra of the relativistic particles. These mainly depend on the value 
of \.{E}, whereas, large values of E$_{el}$ mainly trigger the 
acceleration of particles. 

As seen
from Figure 1 and expression (6), as \.{E} decreases the X-ray luminosity 
of pulsars also decreases. So, in order to have the probability to observe 
X-ray radiation from a pulsar to be high for a fixed value of E$_{el}$, 
the P value of the pulsar must be small (see expression (4)). On the 
other hand, as seen from Figure 1 and Table 1, the L$_x$ values of 
the young pulsars which are connected to SNRs are $\sim$5 orders of 
magnitude larger than the L$_x$ values of old ms pulsars and this is also 
seen from expression (6). This must be the result of small number of 
relativistic and non-relativistic charged particles in the magnetospheres 
of ms pulsars. 

As seen from Figure 2, the difference in the $\tau$ values of very old ms
pulsars ($\tau$$>$10$^9$ yr) and young pulsars is about 5.5 orders of
magnitude (and from expression (5) the difference in the E$_{el}$$^2$
values is also about 5.5 orders of magnitude). The differences in the 
\.{E} values is about 3-3.5 orders and in the P values 
is on average 2.5 orders of magnitude. Considering that the expression 
for E$_{el}$ can only roughly be determined as mentioned above, these 
data show that the relation between \.{E}, E$_{el}$ and P in expression 
(4) is valid.

In order to tear off particles from the surface of neutron star it is 
necessary to have large values of E$_{el}$. The occurence of this process 
must be easier under the existence of hot and extended atmosphere. 
Therefore, the values of mass and age of pulsars must also have important 
roles in the formation of the X-ray emission. 

If the braking index n is constant along all the evolutionary tracks, then 
the age of the pulsar is
\begin{equation}
t = \frac{P}{(n-1)\dot{P}} (1-(\frac{P_0}{P})^{n-1}) = \tau \times   
(1-(\frac{P_0}{P})^{n-1}).
\end{equation}
For the young pulsars P$_0$ (the initial spin period of pulsar) can be 
assumed to be much less than P, so
that, $\tau$ can approximately be equal to the age. For recycled ms
pulsars, the P value can be comparable to the P$_0$ value (the period 
which the radio pulsar is born with), so that $\tau$
can be several times larger than the age. For our purposes such a
difference is not significant since it is enough to know the order of
magnitude of the age which is much longer than the cooling time. 

As the last 4 pulsars displayed in Table 1 with \.{E} values close to 
3$\times$10$^{33}$ erg/s can emit X-ray, all of the single  
old ms pulsars with \.{E}$>$3$\times$10$^{33}$ erg/s must also have X-ray 
radiation. On the other hand, all the 5 old ms pulsars with P$\le$6 ms 
which are located up to $\sim$0.5 kpc (ATNF 2003; Guseinov et al. 2002) 
have been detected in X-ray (Possenti et al. 2002; Becker \& 
Aschenbach 2002). Since the number density of old ms pulsars is very large 
compared to the number density of low mass X-ray binaries, the ages 
of these pulsars must not be much less than $\tau$ in general. As we see, 
although the recycled millisecond pulsars are very old, they still emit 
X-ray because they have large \.{E} values. 

The X-ray luminosities of pulsars together with their PWNe are 
represented in Table 1. Often, such X-ray luminosity values include both 
the X-ray radiation coming from the pulsar and the PWN together, because 
it is difficult to 
separate the emission of the pulsar from the emission of the PWN. 
Therefore, the radiation of single ms pulsars must be smaller compared to 
single young pulsars for equal values of \.{E}. This must be true also 
because of young pulsars having large values of electric field intensity 
and possibly sometimes having smaller masses. Therefore, it is strange 
that ms pulsar J1824-2452 has larger luminosity compared to young pulsar 
J1801-2451 (see Figure 1). Actually, the uncertainties in the luminosity 
values of these ms pulsars may be large. 

In Figure 3, P-\.{P} diagram of all the pulsars in Table 1, which have 
connections with SNRs and/or from which X-ray radiation has been 
observed, is displayed. As seen from this figure and Table 1, pulsars 
Geminga and J0538+2817 which is connected to SNR G180.0-1.7 (Romani \& Ng 
2003) have about one 
order of magnitude larger values of \.{E} compared to old ms pulsars, but 
their luminosity in 2-10 keV band is similar to the luminosity of 
these ms pulsars. The X-ray spectra of the mentioned young pulsars are 
steeper compared to the spectra of ms pulsars as seen from the comparison 
of the luminosities in 0.1-2.4 keV and 2-10 keV bands. The luminosities of 
J0826+2637 and J0953+0755 in 2-10 keV band are smaller than the 
luminosities of old ms pulsars, but their \.{E} values are also smaller 
(see Figure 3). How can we explain such a situation?

Young pulsars have values of electric field intensity about 50 times 
larger than the values of old ms pulsars, they are hotter and they may 
also have 
smaller masses. These facts create conditions for tearing off charged 
particles more easily from neutron stars. On the other hand, the magnetic 
field values of young pulsars are more than 3 orders of magnitude larger 
(the high magnetic field does not let the particles to escape from the 
magnetosphere). 
These and the possible existence of large number of charged particles in 
the magnetospheres and the surroundings of young pulsars must create the 
best conditions for magneto-braking and Coulomb radiation. The small 
values of X-ray radiation of J0538+2817 and Geminga in 2-10 keV band seem 
to contradict this natural discussion. We can 
attempt to get rid of this contradiction if we take into consideration the 
considerably large values of radiation of the mentioned young pulsars in 
0.1-2.4 keV band. The large magnetic field and density of charged 
particles may create conditions under which the energy loss of particles 
in 0.1-2.4 keV band surpasses the energy which is received from 
magnetodipole radiation wave for the further acceleration. 

The existence of large number of accelerated particles and high values of 
magnetic field is well a basis for strong radio radiation. The radio 
luminosities at 1400 MHz of 4 old ms pulsars are about 1-2 order of 
magnitude smaller than the radiation of young pulsars as mentioned above. 
The synchrotron radiation of electrons in unit time is proportional to 
the value of B. As magnetic field of ms pulsars is $\sim$3 orders of 
magnitude smaller than the considered young pulsars, the magnetosphere of 
ms pulsars must contain more than 1 order of magnitude more 
ultrarelativistic particles compared to the magnetosphere of young 
pulsars. This comes from the comparison of the intensity of the radio 
radiation. On the other hand, the maximum radiation at the given 
frequency 
is $\sim$BE$^2$, where E is the electron energy. Therefore, the energy of 
electrons which is responsible for the radiation at 1400 MHz in the case 
of ms pulsars is also large. This shows that the acceleration of charged 
particles takes place in the field of magnetodipole radiation wave and  
number density of the charged particles, which interact with the 
accelerated particles, in the atmosphere of ms pulsars is small. So, 
the cause of the hard X-ray spectra of ms pulsars is understandable. 

\section{Pulsar wind nebulae and pulsars in supernova remnants}
Let us now examine the presence or absence of PWN
around pulsars and the X-ray radiation. All the
pulsars with $\tau$$<$10$^7$ yr represented in Table 1 
are shown in the P-\.{P} diagram in Figure 4 denoting
also the types of the SNRs which some of the pulsars
are connected to. If the type of the SNR is C (composite) or F
(filled-centre), then it means that there is PWN created by the neutron
star. In S (shell) type SNRs there is no observed PWN possibly because it
is very faint. Other pulsars in Table 1 which 
are not connected to SNRs and which have $\tau$$<$10$^7$ yr 
are also displayed in Figure 4. All the pulsars with PWN around them have
L$_x$ (2-10 keV) $>$ 10$^{33}$ erg/s (see Table 1). The cooling radiation 
of these pulsars do not have a considerable role. 

Pulsars J1952+3252, J2337+6151, and J0659+1414, which are connected to 
S-type (in radio band) SNRs, have been observed to radiate in X-ray band. 
Pulsar J1952+3252 has the highest value of L$_x$ (2-10 keV) 
among these pulsars and one can expect to observe PWN around this pulsar 
considering also the small distance and the position of it in the 
Galaxy (see Table 1). On the other hand, the SNR
which this pulsar is connected to (CTB 80) is a well known very old SNR.
Pulsars J2337+6151 and J0659+1414 have also suitable distance values and 
positions to observe possible PWNe around them, but \.{E} and L$_x$ (2-10 
keV) values of these pulsars are so small that it is not so likely that 
PWNe with observable brightness are present around them.

From the analysis of these data, we see that lifetime of the X-ray  
radiation produced by the relativistic particles in the magnetospheres of 
neutron stars is longer than lifetime of PWNe (see Table 1). But
there may be one exception, namely pulsar J0538+2817, which is
connected to SNR G180.0-1.7 (S147). This SNR is S-type in radio band
(Green 2001). Romani \& Ng (2003) claim evidence for a faint nebula around
the pulsar, but in a more recent work (McGowan et al. 2003) no evidence 
has been found of a PWN and the pulsar has been observed to radiate only   
thermally. Also, the positions and ages of the pulsar and the SNR show 
that a physical connection between these two sources is dubious and there 
are no data directly showing evidence for the connection. 

There is another interesting problem: it is known that more energetic
particles have shorter lifetime compared to less energetic particles in a
PWN. So, one may expect to see the diminishing of the X-ray radiation 
of PWN long before the diminishing of its radio radiation, i.e. 
one may expect  
to observe the PWN in radio band for a much longer time compared to
observing the X-ray PWN. We have examined the PWNe both in the radio and
the X-ray bands and we have seen that the lifetime of radio PWN is
comparable to the lifetime of X-ray PWN based on the up-to-date
observational data. So, maybe the ultrarelativistic particles escape from
the PWN or the PWN disrupts shortly after the high-energy partisles in
the PWN which produce the X-ray radiation come to the end of their   
lifetime. Also, the X-ray luminosity of pulsars in 2-10 keV band drops 
down to $\sim$10$^{33}$ erg/s at about the same time when the PWN becomes 
unobservable both in radio and X-ray bands.

From Figure 4 and Table 1 we see that PWN may exist around pulsars with 
\.{E}$>$10$^{35}$ erg/s and $\tau$$<$5$\times$10$^4$ yr. Pulsar  
J1646-4346 in C-type SNR G341.2+0.9 satisfies these criteria, but X-ray 
radiation has not been observed from it because it is located at a 
distance of 6.9 kpc (Guseinov et al. 2003b). X-ray radiation has 
not been observed also from 9 far away pulsars with d$\ge$3 kpc which are 
located close to the Galactic plane in the Galactic central directions 
(see Table 1 and Figure 4) and which satisfy the conditions of large 
value of \.{E} and small value of $\tau$. As seen from the position of 
pulsar J1617-5055 in Figure 4, it must have a connection with SNR and 
must have PWN. This follows also from the large values of X-ray 
luminosity in both considered bands. Lack of observed SNR and PWN 
must be related with the large distance and the position in the Galaxy. 
Ages of the relatively nearby pulsars which are displayed in Figure 4 may 
be smaller than the values of $\tau$. Three of these pulsars are 
connected to S-type SNRs and all of them have small X-ray luminosity.   

\section{Conclusions}
From the analysis of all the available data of single X-ray pulsars, their 
wind nebulae and the SNRs which are connected to some of these sources and 
also from theoretical considerations we have found the conclusions below: 
\\ 
1. The electric field intensity (voltage) of neutron stars is not the main 
physical quantity for the spectra, energy, and intensity of the  
relativistic 
particles which produce the X-ray radiation of single pulsars and their 
wind nebulae. The voltage mainly triggers the acceleration of particles. 
In order to tear off charged particles from the surface of neutron star 
E$_{el}$ must be large. This process occurs more easily in hot and 
extended atmospheres. Therefore, the values of mass and age of pulsars 
also have important roles in the formation of the X-ray radiation. \\
2. The acceleration of relativistic particles mainly takes place in the 
field of the magnetodipole radiation wave. Because of this, the X-ray 
radiation of pulsars and their wind nebulae strongly depend on the value 
of rate of rotational energy loss which is reflected by the spectra of the 
magnetodipole radiation. \\
3. The high magnetic field and particularly the number density of the 
charged particles create conditions under which the energy loss of the 
particles in 0.1-2.4 keV band may surpass the energy received from the 
magnetodipole radiation wave for further acceleration. This must be 
responsible for the steeper X-ray spectra of young pulsars as compared to 
ms pulsars. \\
4. PWN exists only around pulsars with \.{E}$>$10$^{35}$ erg/s and 
$\tau$$<$5$\times$10$^4$ yr. Also, the X-ray luminosity of pulsars in 2-10 
keV band drops down to $\sim$10$^{33}$ erg/s at about the same time when 
the PWN becomes unobservable in both radio and X-ray bands. The lifetime 
of S-type SNRs may exceed the lifetime of F and C-type SNRs. So, it may 
be possible that C-type SNRs can show themselves as S-type after some time 
in their evolution. PWN must be observed around pulsar J1617-5055.

\clearpage
\begin{figure}[t]
\vspace{3cm}
\includegraphics{mass_of_ns_table_GENEL.ps}
\end{figure}

\clearpage
\begin{figure}[t]
\vspace{3cm}
\includegraphics{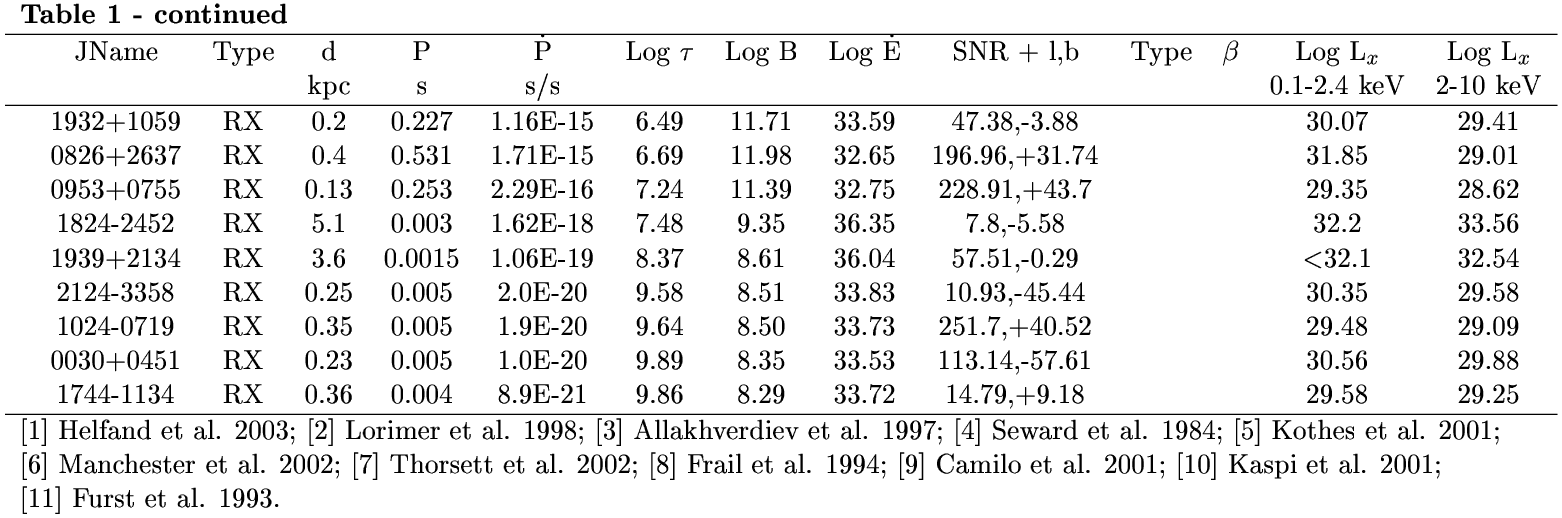}
\end{figure}

\clearpage
{\bf Figure Captions} \\
{\bf Figure 1:} Dependence of the 2-10 keV luminosity on the rate of 
rotational energy loss for pulsars located up to 5 kpc from the Sun 
including also 2 pulsars in Magellanic Clouds which are denotes with '+' 
sign. The single strong X-ray pulsar from which no radio radiation has 
been detected is represented with sign 'star'. Pulsars with 
$\tau$$<$10$^5$ yr are shown with 'X' sign. The old millisecond pulsars 
are displayed with 'square'. \\
{\bf Figure 2:} Dependence of the 2-10 keV luminosity on characteristic 
time for pulsars located up to 5 kpc from the Sun including also 2 
pulsars in Magellanic Clouds. The pulsars are denoted with the same 
symbols as in Figure 1. 
{\bf Figure 3:} Period versus period derivative diagram for the pulsars 
which have been detected in X-ray and/or which are connected to SNRs (see 
Table 1). The sign 'star' represents the pulsars which have genetic 
connections with SNRs and the 'cross' sign denotes the pulsars which are 
not connected to SNRs. \\   
{\bf Figure 4:} Period versus period derivative diagram for all 48 pulsars
represented in Table 1 with $\tau$ $<$ 10$^6$ yr and distance $\le$ 7 kpc
including the 2 pulsars in Magellanic Clouds. 'Circles' represent the 11
pulsars connected to C-type SNRs, 'triangles' show the 9 pulsars connected 
to F-type SNRs, and 'squares' denote the 3 pulsars connected to S-type 
SNRs. 'Plus' signs represent the 14 pulsars which are not connected to 
SNRs and not observed in X-ray. The 11 pulsars which are denoted with 
'stars' have been observed in X-ray but they are not connected to SNRs. \\

\clearpage
\begin{figure}[t]
\vspace{3cm}
\includegraphics{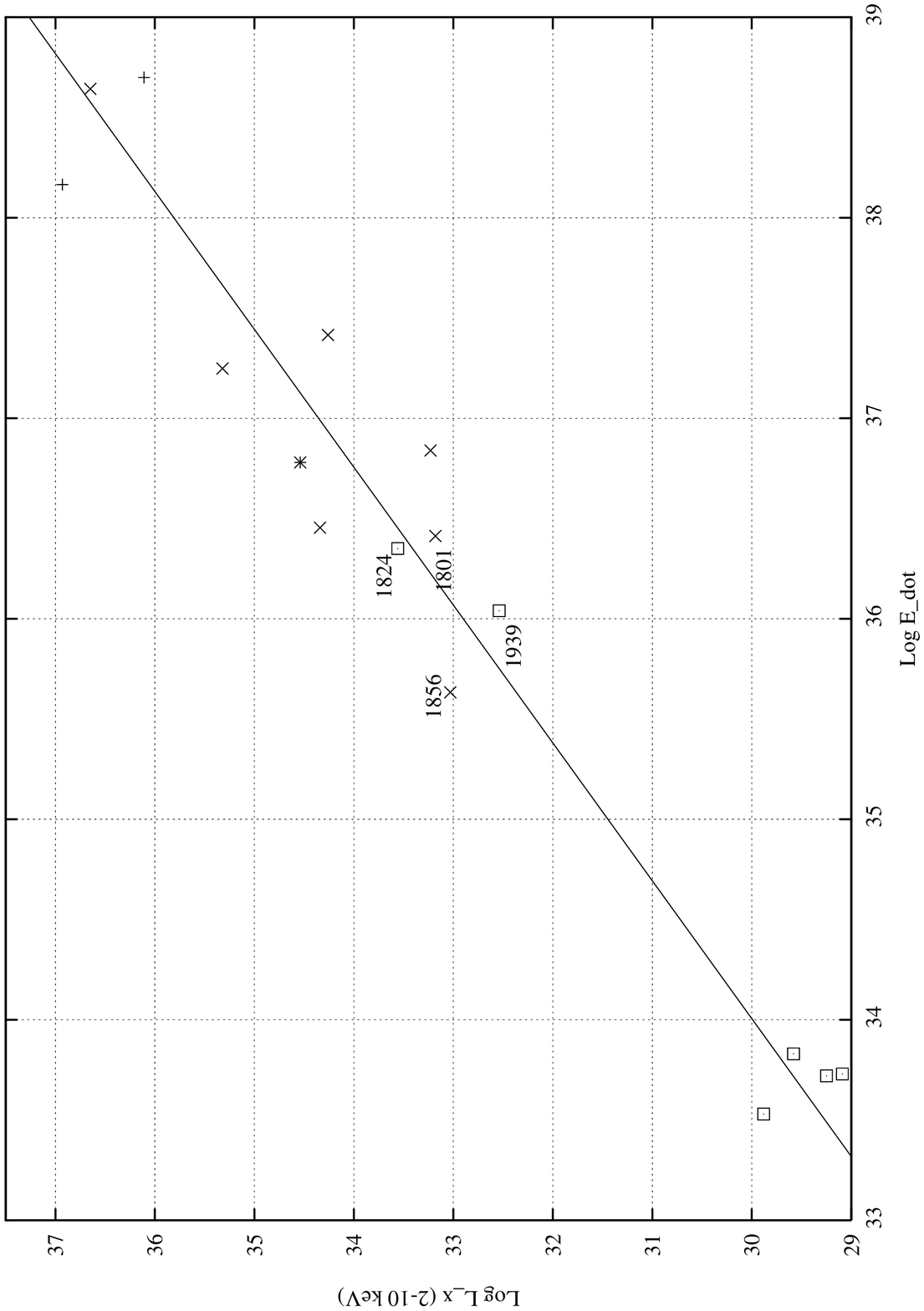}
\end{figure}

\clearpage
\begin{figure}[t]
\vspace{3cm}
\includegraphics{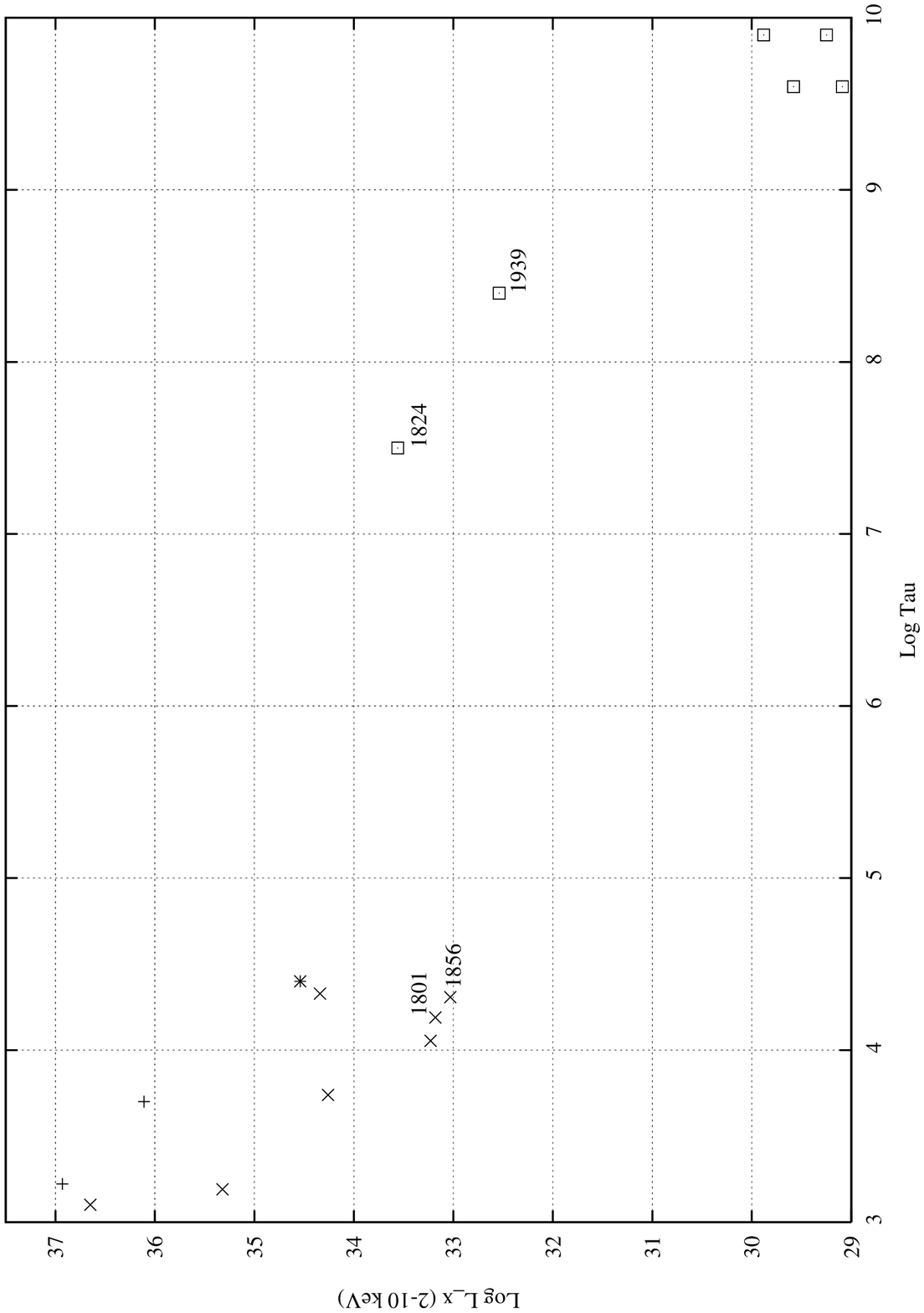}
\end{figure}

\clearpage
\begin{figure}[t]
\vspace{3cm}
\includegraphics{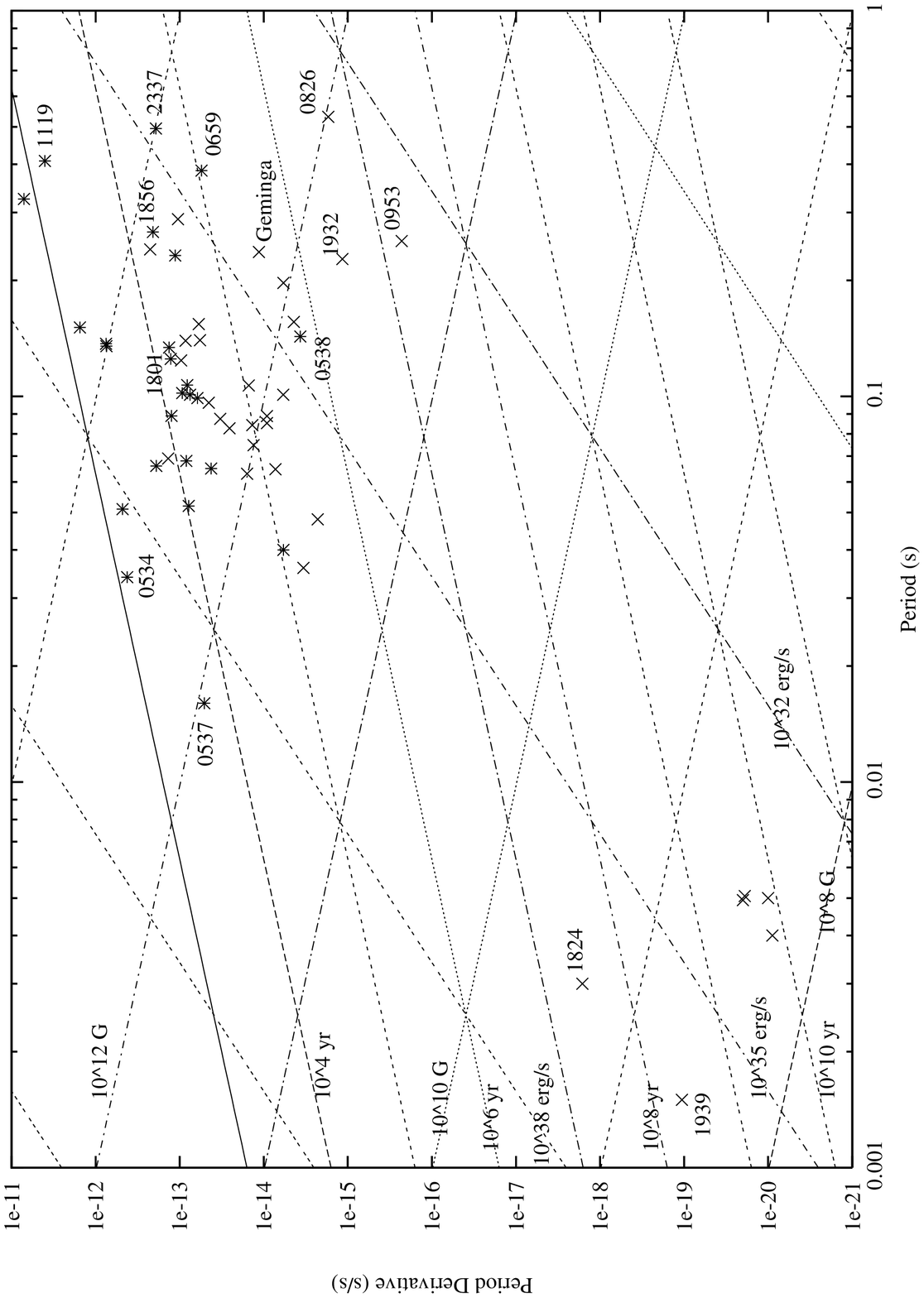}
\end{figure}

\clearpage  
\begin{figure}[t]
\vspace{3cm}
\includegraphics{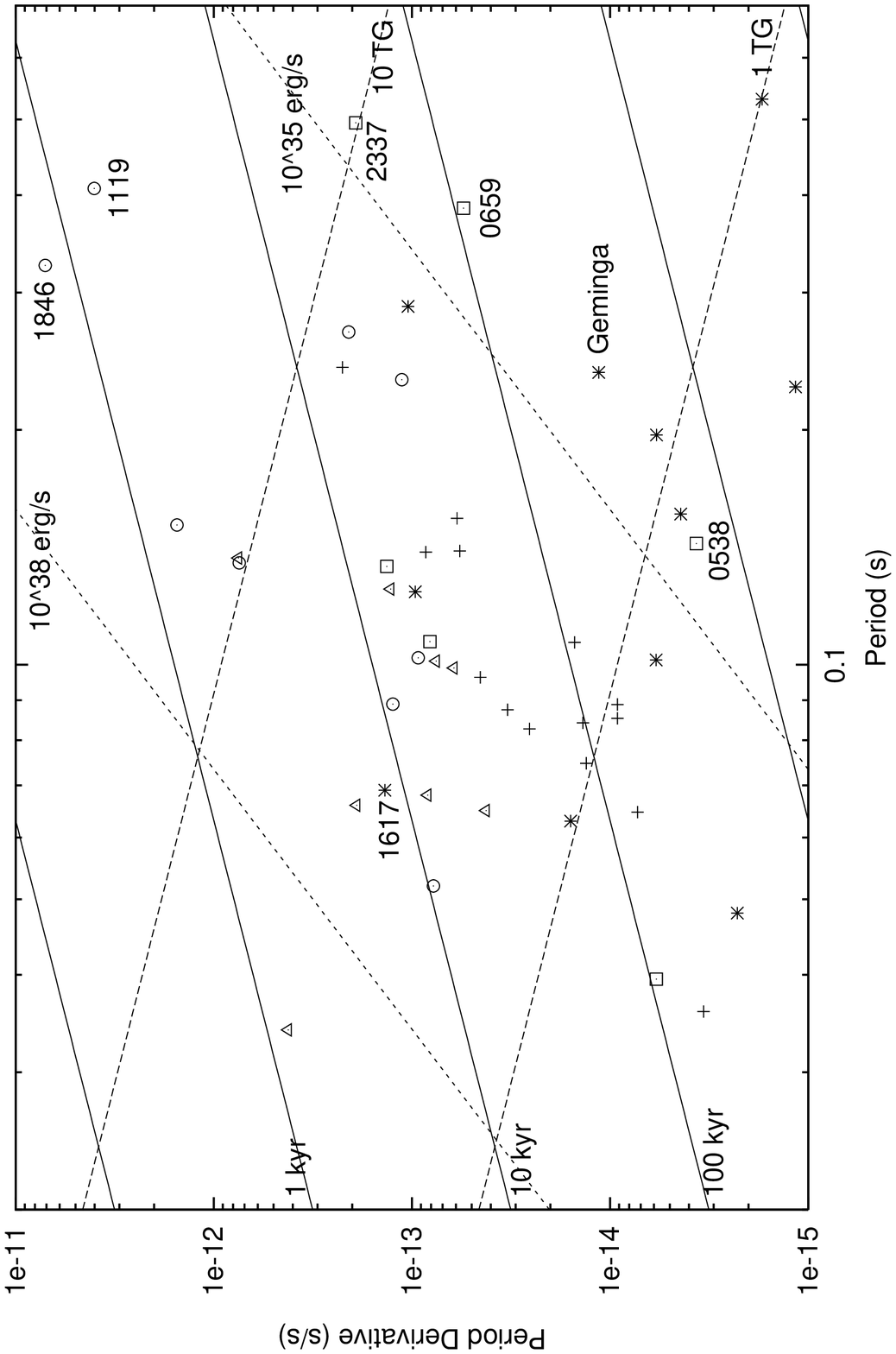}
\end{figure}

\end{document}